\documentclass[12pt]{article}
\usepackage[a4paper, total={6in, 8in},left=20mm,right=22mm]{geometry}
\usepackage[utf8]{inputenc}
\usepackage[english]{babel}
\usepackage{amsmath}
\usepackage{graphicx}
\graphicspath{ {images/} }
\usepackage{float}
\usepackage{amssymb}
\usepackage{caption}
\usepackage{subcaption}
\usepackage{adjustbox}
\usepackage{tabularx}
\usepackage{multicol}
\usepackage{multirow}
\usepackage{slashed}
\usepackage{feynmp-auto}
\usepackage{comment}

\usepackage[tableposition=below]{caption}
\usepackage{longtable}
\usepackage{color}			
\usepackage{verbatim}

\newcommand{\beq}{\begin{eqnarray}}
\newcommand{\eeq}{\end{eqnarray}}

\numberwithin{equation}{section} 
\usepackage[
	colorlinks=true,
	citecolor=red,
	linkcolor=blue,
	urlcolor=blue,
	hypertexnames=false]{hyperref}

\begin{document}
\begin{titlepage}
\begin{flushright}
  {\small
     FERMILAB-PUB-20-256-T\\
\today
}
\end{flushright}
\vspace{1cm}
\begin{center}

	{	
		\LARGE \bf 
		Axion Searches with Two Superconducting Radio-frequency Cavities\\}
	
\end{center}
	\vskip .3cm
	
	\renewcommand*{\thefootnote}{\fnsymbol{footnote}}

\vspace{0.9cm}
\begin{center}
		
		\bf
		Christina Gao\footnote{\tt \scriptsize
		 \href{mailto:yanggao@fnal.gov}{yanggao@fnal.gov},
		 $^\dag$\href{mailto:roni@fnal.gov}{roni@fnal.gov}
		 },
		Roni Harnik$^{\dag}$
\end{center}
	
	\renewcommand{\thefootnote}{\arabic{footnote}}
	\setcounter{footnote}{0}


\begin{center} 

	{\it Theoretical Physics Department, Fermi National Accelerator Laboratory, Batavia, IL, 60510, USA}

\end{center}

\vspace{1cm}

\centerline{\large\bf Abstract}
\begin{quote}
We propose an experimental setup to search for Axion-like particles (ALPs) using two superconducting radio-frequency cavities. In this light-shining-through-wall setup the axion is sourced by two modes with large fields and nonzero $\vec E\cdot \vec B$ in an emitter cavity. In a nearby identical cavity only one of these modes, the spectator, is populated  while the other is a quiet signal mode. Axions can up-convert off the spectator mode into signal photons. We discuss the physics reach of this setup finding potential to explore new ALP parameter space. Enhanced sensitivity can be achieved if high-level modes can be used, thanks to improved phase matching between the excited modes and the generated axion field. We also discuss the potential leakage noise effects and their mitigation, which is aided by $\mathcal{O}$(GHz) separation between the spectator and signal frequencies. 
 
 \end{quote}

\end{titlepage}

\section{Introduction}

The Axion-like particle (ALP) is a natural and well motivated extension of the standard model.  It is a pseudo Nambu-Goldstone boson, arising from the spontaneous symmetry breaking (SSB) of an approximate global $U(1)$, thus naturally light. 
It may be viewed as generalization of the QCD axion, which was originally invented to solve the Strong CP problem~\cite{PhysRevLett.38.1440,PhysRevD.16.1791,PhysRevLett.40.223,PhysRevLett.40.279}, but axion-like particles are well motivated in their own right~\cite{Svrcek:2006yi}. 
The axion's couplings to the Standard Model (SM) fields are commonly suppressed by the SSB scale $f$. In the limit that~$f$ is much bigger than the electroweak scale, which is the scenario we focus on in this work, axions become weakly coupled. 
Since weakly interacting light bosons are sufficiently stable, axion-like particles provide a natural candidate for the dark matter (DM) \cite{Preskill:1982cy, Abbott:1982af, Dine:1982ah}. They are, however, challenging to search for due to their elusive nature~\cite{Graham:2015ouw}.

Searches and constraints on ALPs focus mainly on the axion-two-photon vertex ~$g_{a\gamma}aF\tilde{F}$, where $a$ represents the axion field, $F$ represents a photon field and $g_{a\gamma}\propto1/f$. Some of the strongest constraints on $g_{a\gamma}$ rely on the production of ALPs  in the stellar cores and in supernovae via the Primakoff process. 
For example, the observed lifetimes of the horizontal branch (HB) stars from galactic globular clusters constrains $|g_{a\gamma}|<6.6\times 10^{-11}\textrm{GeV}^{-1} (95\%\textrm{CL})$\footnote{All the limits quoted below will be of $95\%\textrm{CL}$.}~\cite{Ayala:2014pea}, because the emitted ALPs could result in excessive energy losses thus shortened lifetimes of those stars.
In the case of SN 1987A, the duration of the observed neutrino burst places a limit on new sources of energy loss, such as an emission of axions through their coupling to baryons~\cite{Raffelt:1990yz}. In addition,
the lack of $\gamma$-ray signal, which the emitted ALPs can convert into in the galactic $B$-field, places a limit as strong as $|g_{a\gamma}|<5.3\times 10^{-12}\textrm{GeV}^{-1}$ for $m_a\lesssim 4.4\times 10^{-10}$~eV~\cite{Payez:2014xsa}.
Moreover, the direct search for solar axions established a limit of $|g_{a\gamma}|<6.6\times 10^{-11}\textrm{GeV}^{-1} $ for $m_a<0.02$~eV~\cite{Anastassopoulos:2017ftl}, by looking for axion-photon conversion in a static magnetic field.
However, the constraints from the HB star cooling, SN 1987A and the solar axions depend on the astrophysical environment where the ALPs are sourced. These bounds could only be as stringent as how well their corresponding astrophysical environments are understood (e.g.~\cite{Bar:2019ifz} for SN1987A). 
In contrast, laboratory searches have the advantage that both the source and detection of ALPs can be well controlled, therefore they are important complementary probes.

A classic laboratory setup for ALP searches is known as the light-shining-through-walls (LSW). In these experiments, a large number of photons are kept in an enclosed region with a strong constant background magnetic field. We will call this region the emitter or production cavity. In the presence of the $g_{a\gamma}$ interaction,  photons will convert to axions,  and escape the enclosure. A similar strong field region, which we call the receiver or detection cavity, is set up nearby to detect axions that convert back to photons.
The current best limit from LSW is achieved by the OSQAR experiment: $|g_{a\gamma}|<3.5\times 10^{-8}\textrm{GeV}^{-1} $ for $m_a<0.3$~eV~\cite{Ballou:2015cka}. 
Operating at optical frequencies with high finesse cavities, the ALPS experiment took advantage of resonant production and detection~\cite{Hoogeveen:1990vq, Sikivie:2007qm} and achieved a limit~\cite{Ehret:2009sq} comparable to that set by OSQAR, with the prospect of an improvement by a factor $\sim10^3$ in ALPS II~\cite{Bahre:2013ywa}. 
At microwave frequencies, the LSW setup has been implemented by the CROWS experiment and has set a comparable limit~\cite{Betz:2013dza}.

In this work we propose a LSW axion search strategy using superconducting radio-frequency (SRF) cavities in the GHz range. 
SRF cavities can have an exceptionally high quality factor, potentially $Q\gtrsim10^{10}$ (see~\cite{Padamsee_2017} for a review). This can help boost the number of photons in the emitter and also the detectable signal power in the receiver. This is already put to use at the Dark SRF, an ongoing LSW experiment at Fermilab searching for dark photons~\cite{DarkSRF, DarkSRF2}. Dark SRF has demonstrated a large field in the emitter cavity, of order $E_\mathrm{peak}\sim72$~MV/m in its high power run\footnote{Dark SRF quotes the accelerating electric field, $E_\mathrm{acc}\sim 40$~MV/m, which is the average magnitude of the field along the cavity axis. For the elliptical cavities of Dark SRF this field is a factor of $\sim 1.8$ lower than the peak field.}. In addition, Dark SRF demonstrated frequency stabilization of order a Hz with a rigid cavity cage and piezo actuators~\cite{DarkSRF2}.
Large constant magnetic fields, however, could result in flux penetration in the SRF cavity, ruining its high $Q$ property.
To circumvent this problem we will rely on oscillatory cavity modes~\cite{Sikivie:2010fa} that will be excited in both the receiver and the emitter. 

\begin{figure}[t]
     \centering
  \includegraphics[width=12cm]{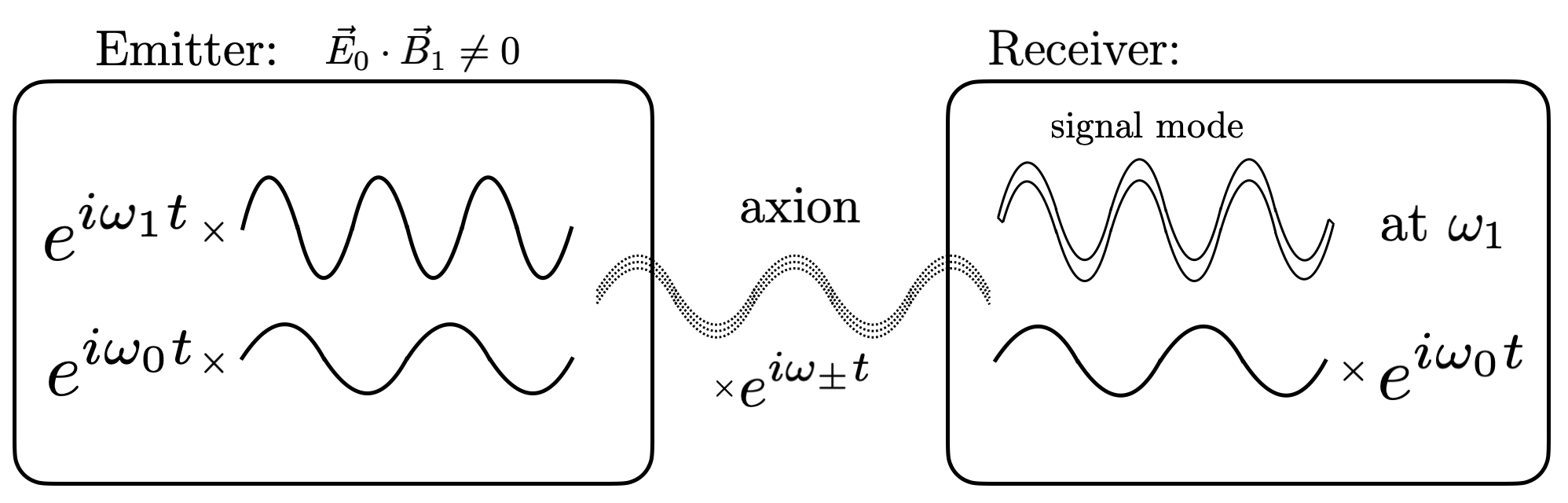}  
\caption{A schematic sketch of the setup.}
\label{fig:setup}
\end{figure}

The setup is shown in Figure~\ref{fig:setup}. The emitter and the receiver are assumed to be two identical cavities placed in close proximity. 
The ALPs are sourced by running two distinct cavity modes at frequencies $\omega_1,\omega_0$ in the emitter cavity. The modes are chosen to have  a non-vanishing $\vec{E}\cdot\vec{B}$ between them. 
The resultant axion field will oscillate with frequencies $\omega_a=\omega_{\pm}\equiv|\omega_0\pm\omega_1|$, and have an amplitude $\sim g_{a\gamma}V_\mathrm{pc}(\vec{E}_1\cdot\vec{B}_0)/r$, where $V_\mathrm{pc}$ represents the volume of the production cavity. 
The receiver cavity is populated with one of the two modes that is responsible for the ALP production, say mode-0. In the presence of this spectator mode the axion has a probability of converting into a photon with a frequency $|\omega_0\pm\omega_a|$.
In particular, the mode $E_1$ can be produced on resonance in the detection cavity and will be our signal.

Several axion searches based on SRF cavities have been proposed~\cite{Sikivie:2010fa, Berlin:2019ahk,Lasenby:2019prg,Janish:2019dpr, Bogorad:2019pbu} and are related to our work. In~\cite{Sikivie:2010fa} and~\cite{Berlin:2019ahk,Lasenby:2019prg} a similar receiver cavity setup was proposed for a resonant search for axion dark matter in the GHz and the kHz range respectively and a broadband DM search was proposed in~\cite{Berlin:2020vrk}. LSW searches test the existence of the ALP as a degree of freedom, without requiring it to make up the dark matter, and are thus complementary. Our LSW proposal shares feature with~\cite{Janish:2019dpr} and~\cite{Bogorad:2019pbu} in which the axion is searched for irrespective of dark matter. Indeed, the axion production discussed here using two modes is similar to these works. In~\cite{Janish:2019dpr} the conversion back to photon occurs in an auxiliary toroidal magnetic field and then gets transferred to be amplified in an empty cavity. The additional toroidal magnet and pick-up loop would need to be of high quality, to avoid introduction of losses. In comparison to this work, our proposal does not require the additional conversion region, simplifying the design.
In~\cite{Bogorad:2019pbu} a single cavity is used and the signal mode is a third mode within the axion-producing cavity, satisfying~$\omega_\mathrm{signal}=2\omega_{0}\pm\omega_1$. Here a major challenge is the mitigation of the nonlinear effects that may lead to leakages of power from a spectator to the signal mode, particularly since the signal mode is at a harmonic of excited modes (as in \cite{Bogorad:2019pbu, Berlin:2020vrk}) or near it (as in ~\cite{Berlin:2019ahk,Lasenby:2019prg}). In our proposal, the large separation, of order GHz, between the pump and signal modes may assist in this noise mitigation. Each of these theoretically proposed methods presents its own challenges and should be studied experimentally.

This paper is organized as follows. In Section~\ref{sec:review} we review the dynamics of electromagnetism coupled to an axion-like particle. In Section~\ref{sec:setup} we discuss our setup and estimate the axion production and signal rate. 
In Section~\ref{sec:reach} we estimate the reach of the setup and discuss some potential backgrounds and in Section~\ref{sec:conclusions} we conclude.

\section{Review of Axion Electrodynamics}\label{sec:review}
We briefly review aspects of axion electrodynamics~\cite{Sikivie:1983ip, Sikivie:2013laa, Beutter:2018xfx,Janish:2019dpr}, the theory that describes the dynamics of coupled axions and photons. 
Consider an axion field $a$ and an electromagnetic field $F_{\mu\nu}$ with the interaction between them:
\beq
\mathcal{L}= -\frac14 (F_{\mu\nu})^2+
\frac12(\partial_{\mu} a)^2-\frac12 m_a^2a^2-\frac{1}{4} g_{a\gamma} a F_{\mu\nu}\tilde{F}^{\mu\nu}.
\eeq
The equations of motion for $\vec{E}$ and $\vec{B}$ are given by the Maxwell equations dressed with axionic terms:
\beq\label{eq:maxwell}
\begin{split}
& \vec{\nabla}\times\vec{E}=-\partial_t\vec{B},&
 \vec{\nabla}\cdot \vec{E} = -g_{a\gamma}\vec{B}\cdot\vec{\nabla}a,
 \\
&\vec{\nabla}\times \vec{B}=\partial_t\vec{E}-g_{a\gamma}(\vec{E}\times\vec{\nabla}a-\vec{B}\partial_ta),
&\vec{\nabla}\cdot\vec{B}=0,
\end{split}
\eeq
where 
$E_i\equiv F_{0i}$, $B^{i}\equiv \frac12 \epsilon^{ijk}F_{jk}$.
Equation~\eqref{eq:maxwell} leads to a wave equation for the (divergence-free) cavity mode $\vec{E}$ \cite{67941},  given by\footnote{The cavity field can in general be decomposed into the of sum a divergence-free and a curl-free component~\cite{67941}. However, only the former is resonantly enhanced in cavities. In a previous version of our paper we did not make this decomposition and hence got an additional term on the right hand side of Equation~\ref{eq:waveeqn}. In most setups this leads to an order one change in our results. We thank Kevin Zhou and Asher Berlin for bringing this to our attention.} 
\begin{equation}\label{eq:waveeqn}
\vec{\nabla}^2 \vec{E}-\partial^2_t\vec{E}=-g_{a\gamma}\partial_t(\vec{E}\times\vec{\nabla}a) + g_{a\gamma}\partial_t(\vec{B}\partial_ta)
.
\end{equation}
Terrestrial experiments looking for axion DM ($a_{DM}$) rely on the term $ g_{a\gamma}\partial_t(\vec{B}\partial_ta_{DM})$, 
since $\vec{\nabla}a_{DM}$ is suppressed by the dark matter virial velocity, of order $10^{-3}$. 
The signal field $\vec{E}$ can then be obtained by solving the wave equation perturbatively in the presence of a spectator field $\vec{B}$ and the oscillating background $a_{DM}$.

Turning to the axion, the $a$ field obeys an equation of motion given by 
\begin{equation}\label{eq:aeom}
\partial^2a+m_a^2a =-g_{a\gamma} \vec{E}\cdot\vec{B}\,.
\end{equation}
We can thus make use of a configuration with $\vec{E}\cdot\vec{B}\ne 0$ as a source of axions in the laboratory. 
Assume that $\vec{E}\cdot\vec{B}\propto e^{i\omega t}$, 
the axion field $a$ at an arbitrary point $\vec{x}$ outside the source enclosing $\vec{y}$ can be written as
\begin{equation}\label{eq:phi}
a(\vec{x},t)=-g_{a\gamma} e^{i\omega t}\int_{V_{source}} d^3\vec{y} \,\,\,\frac{e^{-ik|\vec{x}-\vec{y}|}}{4\pi|\vec{x}-\vec{y}|}\left(\vec{E}\cdot\vec{B}\right)_{\omega},
\end{equation}
where $k=\sqrt{\omega^2-m_a^2}$.  
In summary, Equation~\eqref{eq:phi} allows us to produce axions using  configurations with parallel electric and magnetic fields at a controlled frequency; Equation~\eqref{eq:waveeqn} enables us to detect these axions by letting it interact with a spectator electric or magnetic field. In contrast to axion DM searches, the produced axions have a sizable momentum, i.e.~$\vec{\nabla}a$ is large, so that all three terms on the r.h.s. of Equation~\eqref{eq:waveeqn} can potentially contribute to the signal.

\section{Experimental set up}\label{sec:setup}

For the calculations in this work we consider two identical cylindrical cavities\footnote{SRF cavities used in experiments are often of elliptical shape, but their cavity modes share similar features with cylindrical ones.} with radius $R$ and height $L$. They are arranged to be aligned along their central axes, which is the $z-$axis.
In the emitter, we run two cavity modes with the same peak field value $E_{\rm peak}$ and with a non-zero $\vec{E}\cdot \vec{B}$. In the receiver, we run one cavity mode as the spectator, also with $E_{\rm peak}$. 
The emitter sources an ALP field $a\sim -g_{a\gamma} V\vec{E}\cdot\vec{B}  e^{i\omega_a t}/ (4\pi r) $ based on Equation~\eqref{eq:phi}, where $\omega_a$ equals to the sum or difference of the frequencies of the two cavity modes, $\omega_{\pm}$.
Since $a$ falls off as $1/r$, the second cavity that acts as the detector should be placed in the close vicinity of the first cavity. The choice of the modes will affect the coupling of the axion to both cavities, and thus the sensitivity. 
Below we give details of the production of ALPs in this setup and then proceed to their detection.

\subsection{Production of ALPs}
\begin{figure}[t]
     \centering
  \includegraphics[width=10cm]{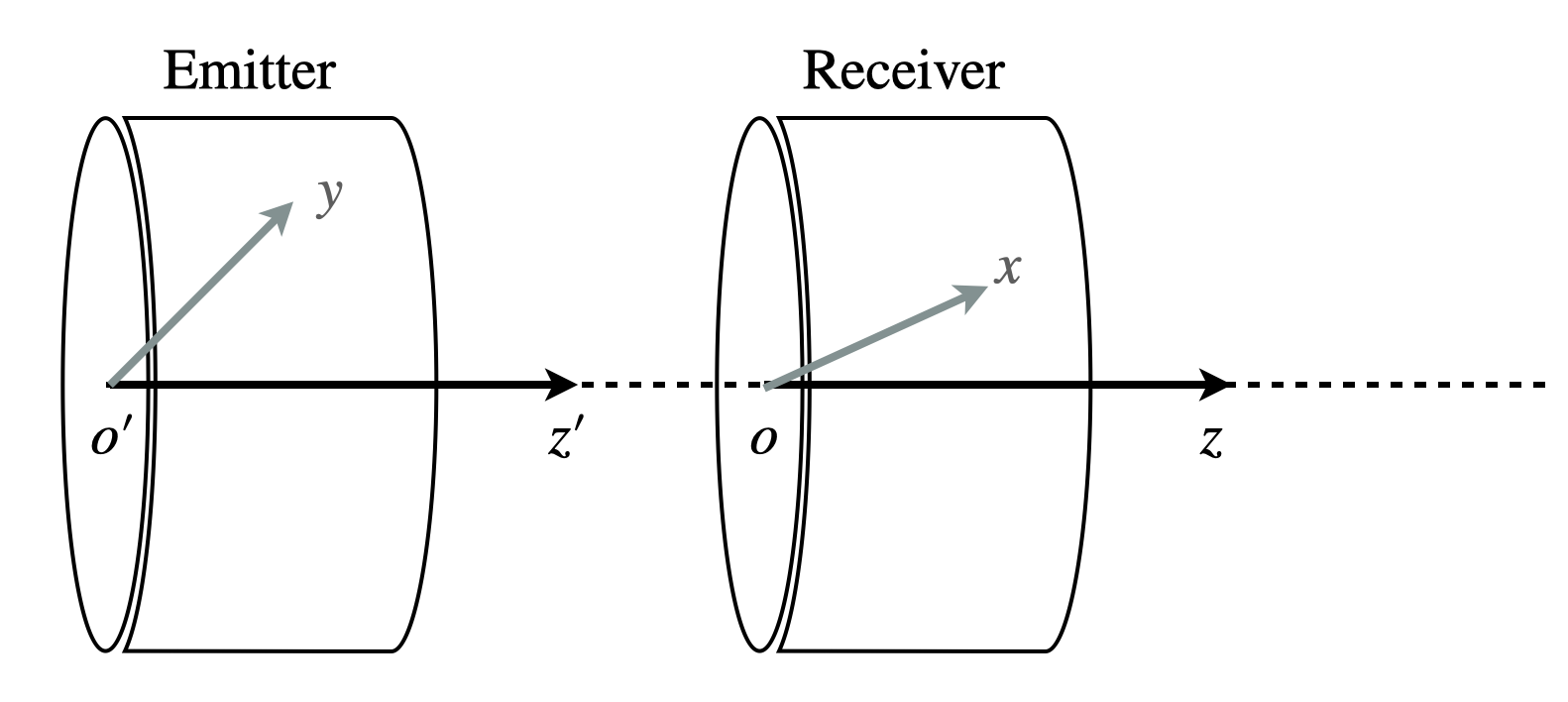}  
\caption{Coordinate setups.}
\label{fig:coord}
\end{figure}

From Equation~\eqref{eq:phi}, at a point $\vec{x}$ outside the production cavity (or emitter) we can decompose the axion field as,
\beq
\begin{split}
a(\vec{x},t)&=a_+(\vec{x},t)+a_-(\vec{x},t),
\\
a_{\pm}(\vec{x},t)&=-g_{a\gamma}e^{-i\omega_{\pm} t}\int _{V_{\rm pc}}d^3y\frac{e^{ik|\vec{x}-\vec{y}|}}{4\pi |\vec{x}-\vec{y}|} \left(\vec{E}\cdot \vec{B}\right)_{\omega_{\pm}},
\end{split}
\eeq
where $k=\sqrt{\omega_{\pm}^2-m_a^2}$. 
Since it is in the receiver that $a$'s amplitude will be of our interest, $\vec{x}$ will be a generic point inside the receiver.
Since the two cylindrical cavities are aligned in their $z-$axes, $a(\vec{x})$ has a rotational symmetry.
Without loss of generality, the distance between a point in the detector $\vec{x}=(r,0,z)$ with respect to the origin at the bottom center of the receiver $\vec{o}$ (c.f. Figure~\ref{fig:coord}), and a point in the emitter $\vec{y}=(r',\theta',z')$ with respect to the origin at the bottom center of the emitter $\vec{o}'$, is given by
\beq
\begin{split}
|\vec{x}-\vec{y}| =& \sqrt{(z+d-z')^2+ r^2+r'^2-2r'r\cos\theta'},
\end{split}
\eeq
where $d$ is the distance between the centers of the cavities, i.e. $d=|\vec{o}-\vec{o}'|$.
The laboratory produced $a$ has both spatial and temporal dependence. In particular,
\beq
\vec{\nabla}_{x}a_{\pm}= g_{a\gamma}e^{-i\omega_{\pm} t}\int_{V} d^3y \,\left(\vec{E}\cdot\vec{B}\right)_{\omega_{\pm}} \frac{e^{ik|\vec{x}-\vec{y}|}(1-ik|\vec{x}-\vec{y}|)}{4\pi|\vec{x}-\vec{y}|^3}
\left(
\begin{array}{c}
r-r'\cos\theta'\\
0\\
d+z-z'
\end{array}
\right).
\eeq

To a have non-vanishing $\vec{E}\cdot \vec{B}$, the two cavity modes must be chosen carefully.
In general, $(\vec{E}\cdot\vec{B})_{\omega_{\pm}}$ will be a linear combination of products of Bessel functions. More details of its form are given in Appendix~\ref{app:vacmodes}. In summary, the emitter produces two axion field $a_{\pm}$, each having its own frequency $\omega_{\pm}$ and amplitude sourced by $(\vec{E}\cdot\vec{B})_{\omega_{\pm}}$.

\subsection{Detection of ALPs}

After ALPs escape the emitter, some of them can convert back to photons in the receiver with a frequency $|\omega_{a}\pm\omega_0|$ in the background of the cavity mode-$0$ . In particular, photons with frequency equal to cavity mode-$1$ will be produced on resonance, and constitute the signal mode. The amplitude of this signal mode-$1$ can be solved via the wave equation \eqref{eq:waveeqn}. The details of this calculation can be found in Appendix~\ref{app:P_sig}. 

Let the spectating mode be
\beq
\vec{B}_{\rm spe}(t,\vec{x})=\vec{B}_0(\vec{x}) b_0(t),\quad \vec{E}_{\rm spe}(t,\vec{x})=\vec{E}_0(\vec{x}) e_0(t),
\eeq
where $b_0(t)\sim ie_0(t)\sim e^{i\omega_0t}$. We define the characteristic amplitude for the spectator as 
\beq
\mathbb{E}_0\,(=\mathbb{B}_0)\equiv (\frac1V \int_V|\vec{E}_0(\vec{x})|^2)^{1/2}\equiv \eta_0  E_{\rm peak},
\eeq 
where $\eta_0 \sim1$.
Similarly, we write the axion field in the factorized form
\beq
a(\vec{x},t)\equiv \mathbf{a}(\vec{x})f(t),\,{\rm where}\, f(t)\sim e^{-i\omega_{a} t}.
\eeq
Let the signal mode be 
\beq
\vec{E}_{\rm sig}(t,\vec{x})=\vec{E}_1(\vec{x})e_1(t), \quad e_1(t)\sim e^{i\omega_1t},
\eeq
where $\vec{E}_1$ is the amplitude we are looking for.

Using the characteristic amplitude for the signal, 
\beq\label{eq:E1def}
\mathbb{E}_1\equiv (\frac1V \int_V|\vec{E}_1(\vec{x})|^2)^{1/2},
\eeq
the solution to Equation~\eqref{eq:waveeqn} is
\beq
\label{eq:sigE}
\mathbb{E}_1\tilde{e}_1(\omega)=\frac{-i\omega g_{a\gamma}\mathbb{E}_0}{\omega^2-\omega_1^2-i\omega\omega_1/Q_1}
\times \int \frac{d\omega'}{2\pi}\tilde{e}_0(\omega-\omega')\tilde{f}(\omega')(\alpha +\beta\omega'
),
\eeq
\\
where
\beq
\begin{split}\label{eq:abc}
&\alpha\equiv\frac{\int_V\vec{E}_1^*\cdot(\vec{E}_0\times\vec{\nabla}\mathbf{a} )}{\sqrt{\int_V|\vec{E}_1|^2}\sqrt{\int_V|\vec{E}_0|^2}},\quad
\beta\equiv\frac{\int_V\vec{E}_1^*\cdot(\vec{B}_0\mathbf{a} )}{\sqrt{\int_V|\vec{E}_1|^2}\sqrt{\int_V|\vec{B}_0|^2}}
.
\end{split}
\eeq
$\tilde{f},\tilde{e}_0,\tilde{e}_1$ are the Fourier transforms of $f(t),e_0(t),e_1(t)$, respectively.

\begin{figure}
     \centering
     \includegraphics[trim =40 0 0 0, scale=0.5]{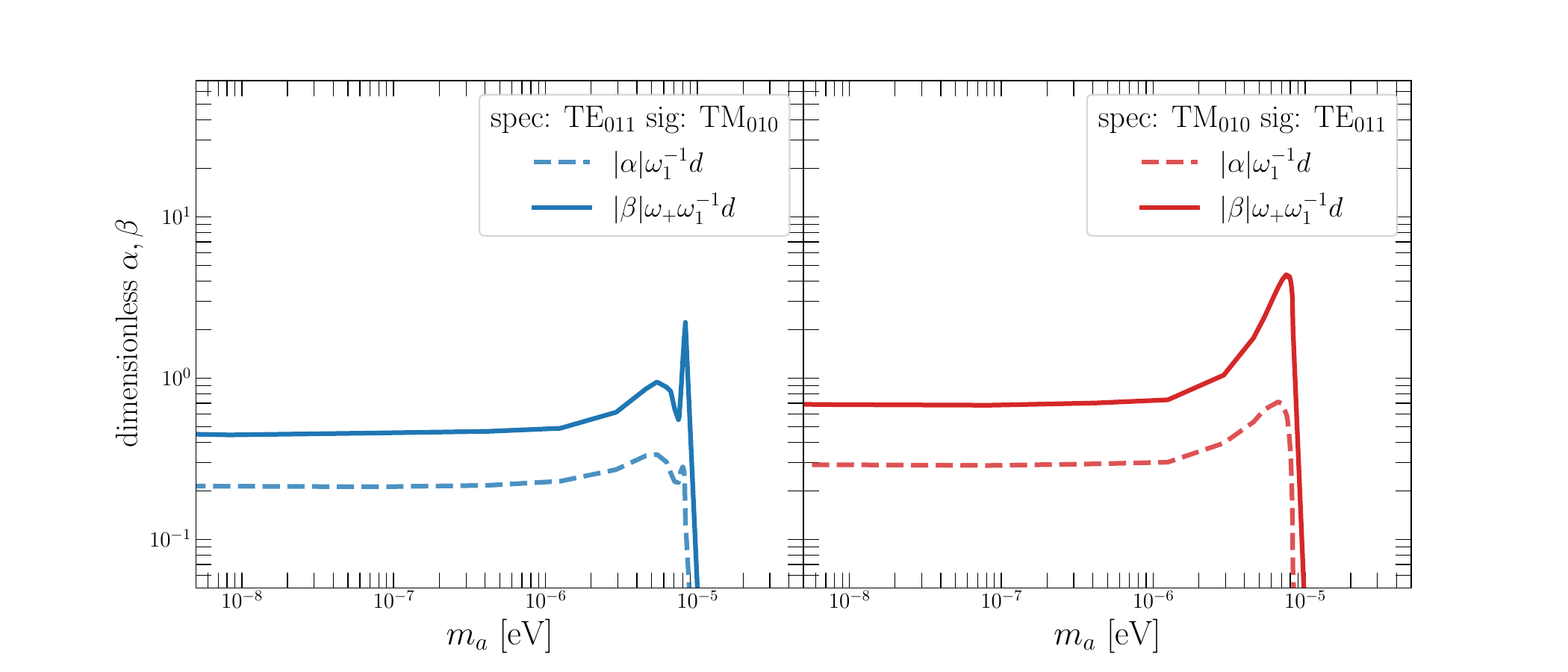}
\caption{Comparisons of $\alpha$ and $\beta$ given by Equation~\eqref{eq:abc}, assuming $g_{a\gamma}=5\times10^{-11}{\rm GeV}^{-1}, E_{\rm peak}=80\rm MV/m$. $\omega_{+}=\omega_{\rm TE}^{011}+\omega_{\rm TM}^{010}$, $\omega_1$ is the frequency of the signal mode for each detector.
Production and detection cavities are cylindrical with $R=L=0.1766\rm m$, and are aligned in $z$ with $(0.1\rm m+L)$ between their centers. 
}
\label{fig:abc}
\end{figure}

Figure~\ref{fig:abc} compares $\alpha$ and $\beta$
for two setups, in which both emitters use a transverse electric (TE) mode and a transverse magnetic (TM) mode to produce axions. The setup on the left (blue) runs the TE mode as the spectator and looks for the TM mode in the detector, whereas the setup on the right (red) does the opposite. 
It is clear that the main contribution to the signal in both setups are from $\beta$, which doesn't depend on the gradient of the axion field.
The sharp cutoff at $\sim10^{-5}$ eV is precisely where $\omega_+$ sits. Axions with a mass bigger than $\omega_+$ cannot be produced on shell, thus resulting in a rapid drop in signal. The quick rise in signal very close to this threshold is due to the phase matching between the massive axion field and the low-lying cavity modes.

Now the signal power can be easily computed. The steady state average power output in the receiver can be expressed as
\beq
\begin{split}\label{eq:P_sig}
P_{\rm sig}=&\frac{\omega_1}{Q_1}\int_{V}|\vec{E}_1(\vec{x})|^2\langle |e_1(t)^2|\rangle ,
\end{split}
\eeq
where
$Q_1$ is the quality factor for $\omega_1$. 
Using Equations~\eqref{eq:E1def} and \eqref{eq:sigE}, we obtain
\begin{equation}
\begin{split}
P_{\rm sig}=&\frac1{16\pi^2}\frac{\omega_1V g_{a\gamma}^2\mathbb{E}_0^2}{Q_1}\int d\omega \frac{\omega^2}{(\omega^2-\omega_1^2)^2+\omega^2\omega_1^2/Q_1^2}\times\\
&\left\{
S_{e_0}(\omega+\omega_{a})|\alpha-\beta\omega_{a}
|^2
+S_{e_0}(\omega-\omega_{a})|\alpha+\beta\omega_{a}
|^2
\right\},
\end{split}
\end{equation}
where $S_{e_0}$ takes the form
\beq
S_{e_0}(\omega)=\pi^2(\delta(\omega+\omega_0)+\delta(\omega-\omega_0)),
\eeq
and can be interpreted as the power spectral density. 
Using the frequency matching condition $\omega_1\pm\omega_0=\omega_{a\pm}$,
\begin{equation}
\begin{split}
P_{\rm sig}
=&\frac 18 V\mathbb{E}_0^2 g_{a\gamma}^2\frac{Q_1}{\omega_1}\left(
|\alpha|^2+\left|\beta (\omega_1\pm\omega_0)
\right|^2
\right).
\end{split}
\end{equation}

\subsubsection{Accounting for $\omega_a=\omega_{\pm}$}

Recall that there are two frequencies of $a$ produced from $\vec{E}\cdot \vec{B}$ in the production cavity, hence $
a=a_++a_-\equiv \mathbf{a}_+(\vec{x})f_+(t)+\mathbf{a}_-(\vec{x})f_-(t)
$. Therefore, the solution to Equation~\eqref{eq:waveeqn} becomes
\beq
\begin{split}
\mathbb{E}_1&\tilde{e}_1(\omega)
=\frac{-i\omega g_{a\gamma}\mathbb{E}_0}{\omega^2-\omega_1^2-i\omega\omega_1/Q_1}
\times \int \frac{d\omega'}{2\pi}\tilde{e}_0(\omega-\omega')\\
&\left\{
\tilde{f}_+(\omega')(\alpha_+ +\beta_+\omega'
)+
\tilde{f}_-(\omega')(\alpha_- +\beta_-\omega'
)
\right\},
\end{split}
\eeq
where $\alpha_{\pm},\beta_{\pm}
$ 
are Equation~\eqref{eq:abc} with $\mathbf{a}$ replaced by $\mathbf{a}_{\pm}$.
Since $\tilde{f}_{\pm}(\omega)\sim\delta(\omega-\omega_{\pm})$, $\langle \tilde{f}_+(\omega)\tilde{f}_-(\omega)\rangle=0$. The signal power including contributions from both frequencies is given by
\beq\label{eq:P_sig_pm}
\begin{split}
P_{\rm sig}=P_{\rm sig}^{+}+P_{\rm sig}^{-},\quad
{\rm where}\,
P^{\pm}_{\rm sig}
=\frac{Q_1}{8 \omega_1}V g_{a\gamma}^2\mathbb{E}_0^2\times
\Big(
\left|\alpha_{\pm}\right|^2+\left|(\omega_1\pm\omega_0)\beta_{\pm}
\right|^2\Big).
\end{split}
\eeq

\begin{figure}
\centering
     \includegraphics[scale=0.5]{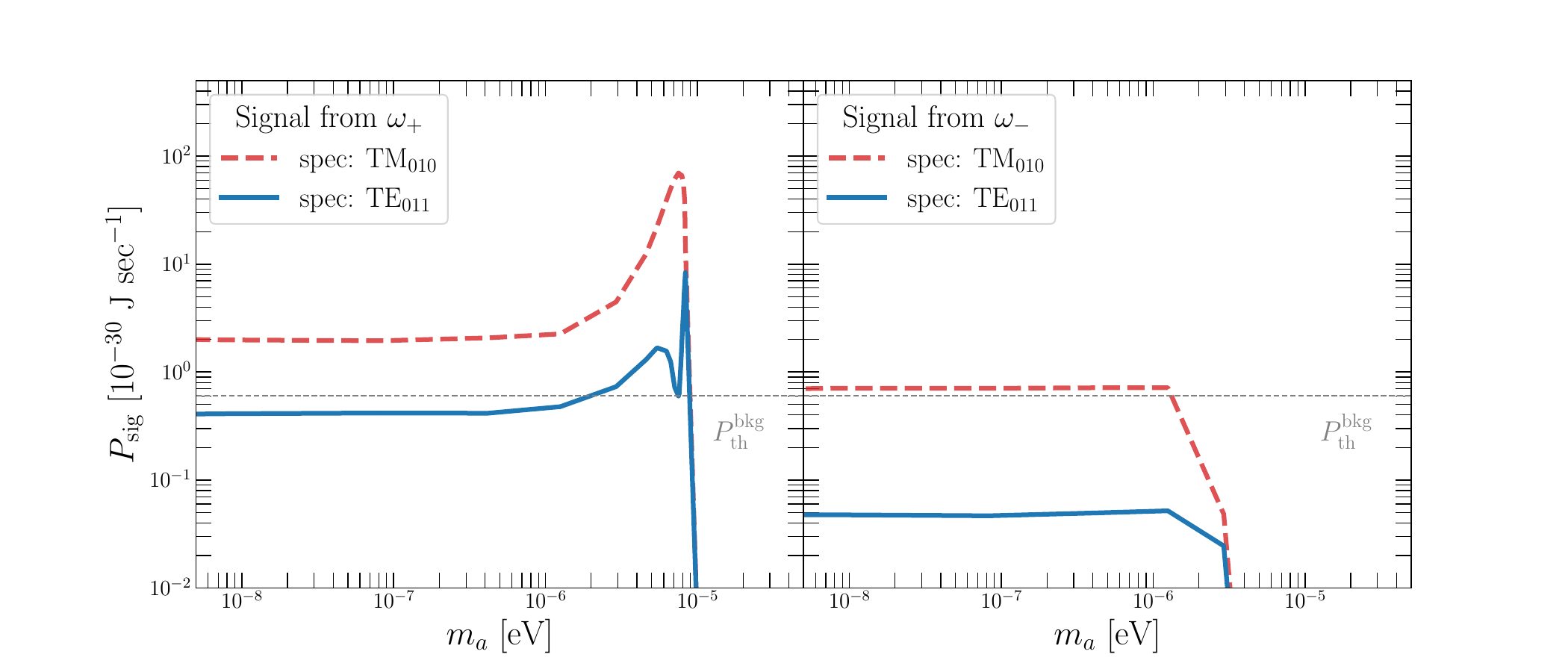}
\caption{Comparisons of signal power, assuming $g_{a\gamma}=5\times10^{-11}\rm GeV^{-1}$, $ E_{\rm peak}=80\rm MV/m$, $ Q_1=10^{10}$. $\omega_{\pm}=\omega_{\rm TE}^{011}\pm \omega_{\rm TM}^{010}$.
The red and blue lines correspond to the two setups in Figure~\ref{fig:abc}. Also shown is the thermal background (grey) equal to $k_BT \Delta\omega_1$ assuming the system is cooled down to $1.4$~K and $\Delta\omega_1={\rm year}^{-1}$. 
}
\label{fig:power}
\end{figure}

Figure~\ref{fig:power} shows the signal power as function of axion mass. The left (right) panel compares $P_{\rm sig}^{+}$ ($P_{\rm sig}^{-}$) for the two setups with the spectator and signal mode switched. It is clear that $P_{\rm sig}^{+}$ is always the dominating contribution to the signal power.

\section{Physics Reach}\label{sec:reach}
To explore the physics reach of the experimental setup above, one needs to compute its signal-to-noise ratio (SNR). Given an integration time $t_{\rm int}$, the SNR is approximately given by \cite{Chaudhuri:2018rqn}:
\begin{equation}
\mbox{SNR}=\frac{P_{\rm sig}}{P_{\rm noise}}\sqrt{t_{\rm int}\Delta\omega_1}=\frac{P_{\rm sig}}{k_BT\Delta\omega_1}\sqrt {t_{\rm int}\Delta\omega_1},
\end{equation}
where $\Delta\omega_1$ is traditionally chosen as $\omega_1/Q_1$ but can be as small as $t^{-1}_{\rm int}$ \cite{Janish:2019dpr}.
We have assumed that the dominant noise is from the thermal noise. Other sources of noise will be discussed in the next subsection.

\begin{figure}[h]

     \includegraphics[trim=20 0 0 0, scale=0.5]{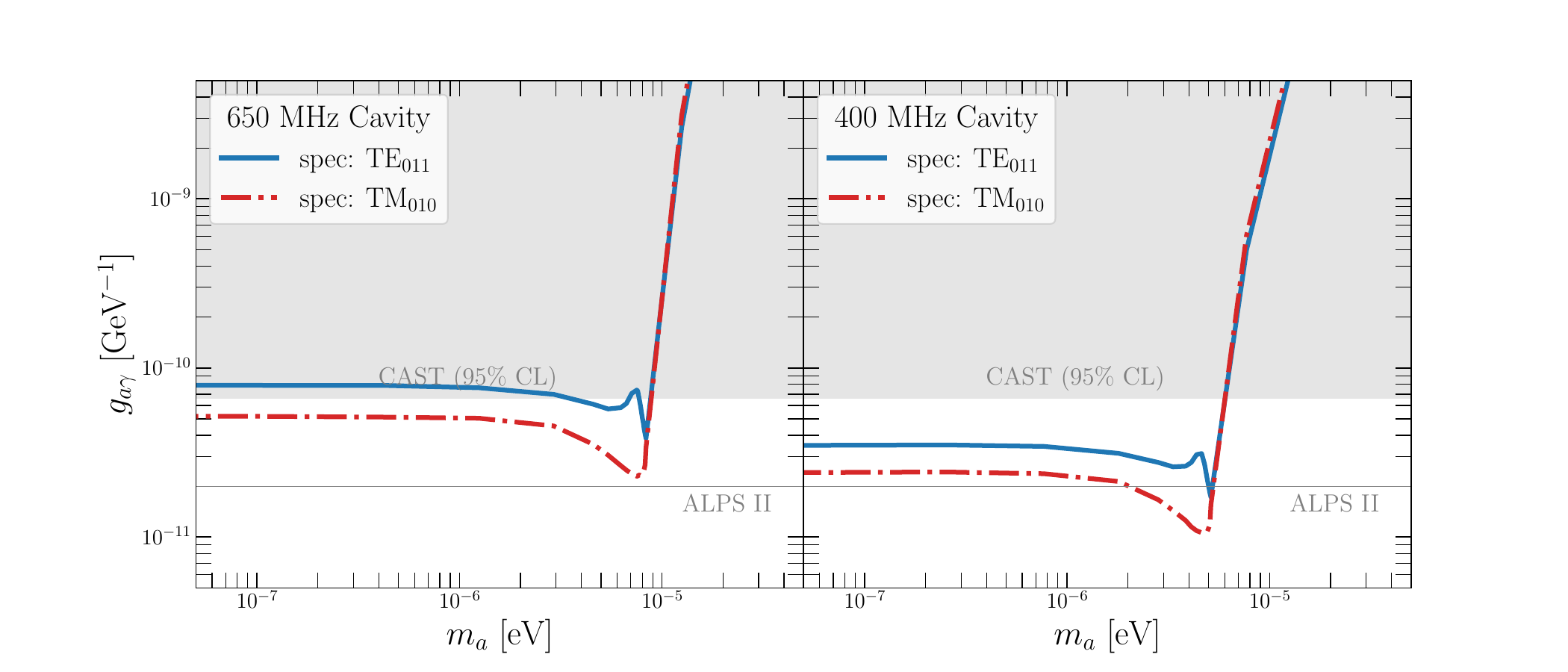}
\caption{SNR $=5$ contours in the plane of $g_{a\gamma}$ vs $m_a$, assuming $ E_{\rm peak}=80\rm MV/m$, $ Q_1=10^{10}$, $T=1.4\rm K$, $\Delta\omega_1=t_{\rm int}^{-1}$, $t_{\rm int}=1\rm year$.
The left (right) panel assumes cylindrical cavities with $R=L=0.1766\rm m\,(0.287\rm m$), aligned in $z$ with $(0.1{\rm m}+L)$ between their centers.
The red and blue contours correspond to the two setups in Figure~\ref{fig:abc}.
Also shown are the limits that are established by CAST and projected by ALPS II.
}
\label{fig:tevstm}
\end{figure}

We may get an order of magnitude estimate of SNR in the limit that $m_{a}\ll \omega_{a}$ and that the separation between the emitter and the detector is much larger than the cavity size. From Equation~\eqref{eq:phi}, the spatial part of $a$ and its gradient are approximately given by
\beq
a(\vec{x})\sim \frac{ \eta_{01} g_{a\gamma} V E
^2_{\rm peak}}{4\pi d},\quad 
\vec{\nabla}a(\vec{x})\sim \frac{ \omega_{a} \eta_{01} g_{a\gamma}V E
^2_{\rm peak}}{4\pi d} \hat{z},
\eeq
where $\eta_{01}< 1$, characterizing the geometric overlap between the two modes that are responsible for the ALP production in the emitter. From Equation~\eqref{eq:sigE},
\beq
\frac{\alpha}{\omega_{a} }\sim \beta
\sim\frac{\eta^2_{01} g_{a\gamma}V E^2_{\rm peak}}{4\pi d},
\eeq
{where we assumed that the geometric overlap between the generated axion field, the spectator and signal modes in the receiver is also roughly given by $\eta_{01}$.
Since $\omega_{a}=\omega_1\mp\omega_0$, $\omega_{a}\sim \omega_1\sim\omega_0\sim V^{-1/3}$. Taking $\Delta\omega_1$ to be as small as $ t_{\rm int}^{-1}$, 
parametrically,
\beq
\begin{split}
\mbox{SNR}\sim &\frac{Q_1}{8\omega_1}V g_{a\gamma}^2\mathbb{E}^2_0
\left(\omega_{a}\frac{\eta^2_{01} gV E^2_{\rm peak}}{4\pi d}
\right)^2  \frac 1Tt_{\rm int}
\\
\sim &
5\,
\left(\frac{Q_1}{10^{10}}\right)
\left(\frac{V}{(0.2\rm m)^3}\right)^3 
\left(\frac {g_{a\gamma}\, \rm GeV }{5\times10^{-11} }\right)^4 \left(\frac{E_{\rm peak}}{80\rm MV/m}\right)^6\\
&\quad\times
\left(\frac{ 0.4\rm m}d\right)^2 \left(\frac{ \omega_a}{ \rm GHz}\right)\left(\frac { t_{\rm int}} {1\rm year}\right)\left(\frac {1.4\rm K} T\right)
\end{split}
\eeq
where 
$\eta_{01}\sim 0.5$, and the volume of the cavity is approximately $(0.2\rm m)^3$ which yields modes with frequencies of $\mathcal{O}(\rm GHz)$. 
We further assume that the GHz cavities of $Q\sim 10^{10}$ are cooled down to 1.4K. The benchmark value we take for the production modes in the emitter and the spectator mode in the receiver are all  $E_{\rm peak}=80\rm MV/m$ which has been demonstrated in non-pulsed cavity tests~\cite{Grassellino:2017bod,Bafia} and during the high power run of Dark SRF~\cite{DarkSRF2} \footnote{We again note that $E_{\rm peak}$ is not the same as $E_{\rm acc}$, which is commonly used in the cavity literature}. With an integration time of one year and these benchmarks, an axion-photon coupling as small as $g_{a\gamma}\sim 5\times10^{-11}\rm GeV^{-1}$ can be probed in the limit of vanishing ALP mass.

Requiring SNR $=5$ yields a limit in the plane of axion-photon coupling $g_{a\gamma}$ versus the axion mass $m_a$ as shown in Figure~\ref{fig:tevstm}. 
It is clear that in our setup choosing TM mode as the spectator yields a better sensitivity. 

It is interesting to compare the reach of our setup to other recent SRF-based proposals for (non-dark matter) axion searches~\cite{Janish:2019dpr, Bogorad:2019pbu}. Like ours in both works axions are produced by exciting two pump modes in an emitter cavity. The former proposal has a conversion region with a static magnetic field which is connected to an empty receiver cavity. The later combines the emitter and receiver, looking for transfer of power from the two pump modes to a third mode which is tuned to be a harmonic. The reach curves shown in these works are somewhat stronger, but this is due to differences in assumptions. For example, the authors of both~\cite{Janish:2019dpr} and~\cite{Bogorad:2019pbu} chose to present limits for receiver $Q$'s of $10^{10}$ and $10^{12}$, whereas we only show a benchmark of $10^{10}$, noting that the limit scales as $Q^{1/4}$. Other differences include an assumed temperature of $0.1$~K in~\cite{Janish:2019dpr}, compared to 1.4~K assumed here, an integration time of two weeks 
as well as a larger cavity volume assumed in~\cite{Bogorad:2019pbu}. When making similar assumptions to phase 2 of~\cite{Bogorad:2019pbu}, we find a reach of $g_{a\gamma}\sim 2\times 10^{-12}$ for low axion masses. The setup in~\cite{Janish:2019dpr} benefits from a stronger magnetic field that is achievable in a static field, as compared to an RF mode, but this comes at the cost of a somewhat reduced form factor (the result of the spatial integrals in the rate calculations). Summarizing this comparison, we find that the various proposed setups have similar reaches for similar assumptions. The important differences lie in the complexity of the experimental setup, as well as the presence of non-thermal sources of backgrounds, which we discuss in Section~\ref{sec:leak}.

\subsection{Excited cavity modes}
In this subsection we investigate the use of excited cavity modes for emission and detection of axions. We allow the signal mode to vary in $\ell$, i.e. TE$_{01\ell}$, but with the spectating mode fixed to be TM$_{010}$. In this case the corresponding production modes become TM$_{010}$ and TE$_{01\ell}$. Figure~\ref{fig:limit} compares the limits that can be achieved by four different $\ell$s. As $\ell$ is increased, not only can we gain access to heavier ALPs,  but also achieve a better sensitivity for light ALPs. 
The improvement at the large axion mass is due to the higher frequency of the high~$\ell$ modes, which allow for a larger kinematic reach. We note however the wiggles near thershold. These are the result of destructive interference for axion masses in which an integer number of axion wavelengths fit in the receiver cavity.
\begin{figure}[h]
     \centering
     \includegraphics[trim=20 0 0 0, scale=0.5]{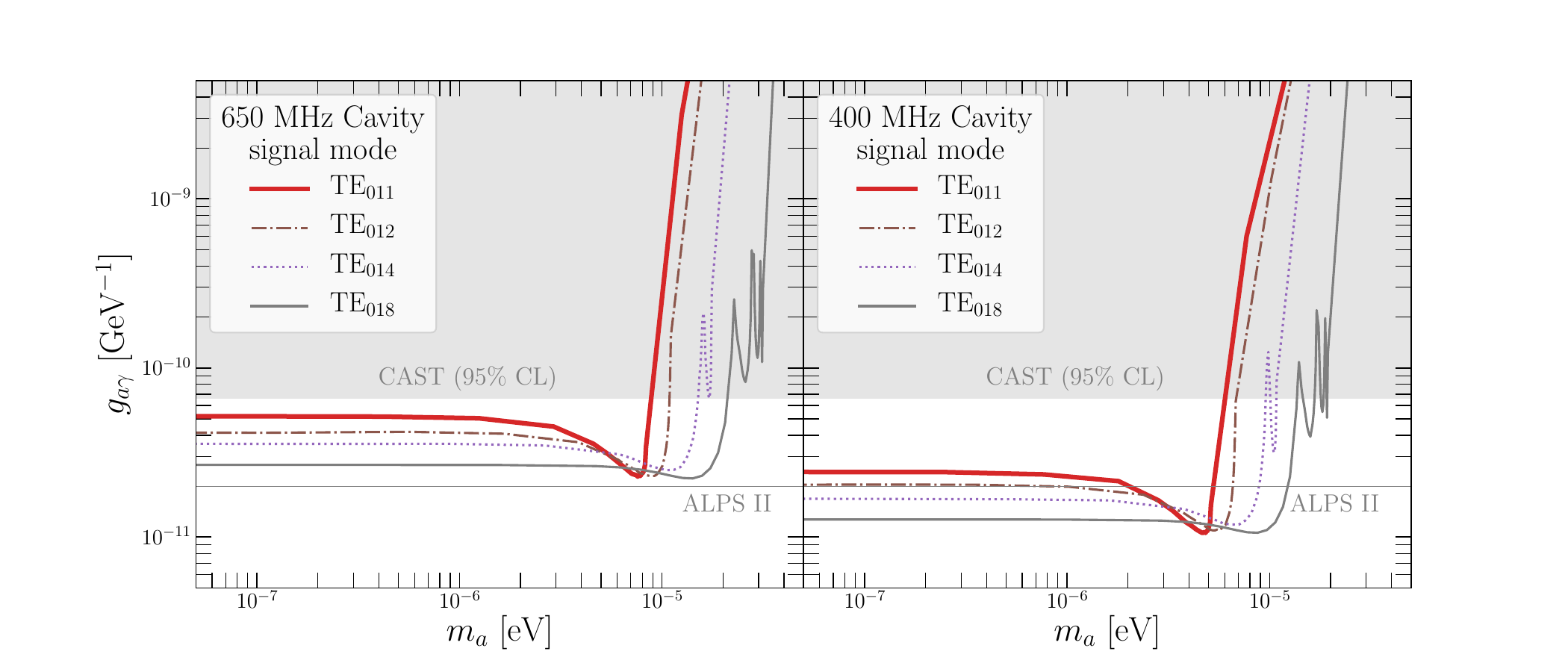}
\caption{Sensitivity curves (SNR $=5$) in the plane of $g_{a\gamma}$ vs $m_a$ assuming different signal modes with a fixed spectating mode TM$_{010}$. We use $E_{\rm peak}=80\rm MV/m$, $ Q_1=10^{10}$, $T=1.4\rm K$, $\Delta\omega_1=t_{\rm int}^{-1}$, $t_{\rm int}=1\rm year$.
The left (right) panel assumes SRF cavities that are cylindrical with $R=L=0.1766\rm m\,(0.287\rm m$), and are aligned in $z$ with $(0.1\rm m+L)$ between their centers.
We see that using higher modes improve sensitivity both at high ALP mass (due to higher kinematic reach) and at low masses (due to improved phase matching between a relativistic axion and excited modes).}
\label{fig:limit}
\end{figure}

The improvement at the low axion mass is due to an improvement in the phase matching between the axion and the cavity mode. For example, in the massless limit, the axion exhibits a relativistic dispersion relation $\omega_a\sim |\vec k_a|$. The overlap integral of the signal mode and the axion will be largest when the signal mode also exhibits the same relation of frequency and wavelength. 
Recalling that excited modes have shorter wavelengths, thus less overlaps with the boundaries of the cavity, it is clear that they will have a dispersion relation that is closer to that of a free massless photon and an enhanced coupling.

We thus conclude that using excited cavity modes can improve the reach of axion searches. We expect that this improvement will be present in other setups, such as the Dark SRF dark photon search~\cite{DarkSRF, DarkSRF2}.  We note, however, that experimentally establishing a high $\ell$ mode with a high quality could be challenging. Since at high frequencies there exist numerous cavity modes neighboring the desired mode, 
it may be difficult both to generate and look for the particular mode we want. This presents another interesting experimental challenge.

\subsection{Leakage Backgrounds}\label{sec:leak}

In the discussion so far we have assumed that the only important background is from thermal fluctuations. In order to reach this level of sensitivity one must mitigate other sources of background. 
Several sources of background, including mechanical noise, oscillator phase noise, fields emission have been considered in \cite{Berlin:2019ahk}, in which the frequency difference of the spectator and signal modes is of order 1-100 kHz. In the region of interest here, splittings of order a GHz, these sources of background can be extrapolated to be subdominant to the thermal noise. In this section we will consider the sources of non-thermal noise which are most likely to be a concern in our setup: leakage directly from the driving source, and leakage due to nonlinear material effect. We point out that both of these noise sources can be mitigated by using a very narrow pump source. In addition they can be further suppressed by optimizing the cavity geometry, and further material science techniques to reduce nonlinearities. 
Particularly with regard to nonlinear material effect,
the goal of the discussion here is not to speculate on the size of this noise source, but rather to motivate an experimental exploration. We note that, as opposed to dark matter searches, the signal in a LSW experiment may be turned off by deactivating or detuning the emitter in order to characterize the noise in the receiver cavity.
 
A particular worry of our multi-mode setup is that the signal mode lives in the same cavity as a spectator mode which is being driven to high occupancy. A small leakage of power either from the driving source or from the spectator mode to the signal mode can easily dominate over the thermal background. 
To put this in perspective, an excited mode with $E_\mathrm{peak}\sim 80$~MV/m at GHz frequencies has roughly $10^{26}$  photons. The thermal background at these frequencies is of order a few thousand photons. 
These challenges were already identified in~\cite{Sikivie:2010fa} and in this subsection we will discuss ways to overcome them. 
Since the signal mode at temperatures of a few Kelvin is still in the classical regime of many photons, we can address this problem classically and ask what the strength of the field is at a small window around the signal frequency $\omega_1$. 
As alluded to above, there are two distinct potential sources of leakage, (\emph{a}) the source which is used to drive the spectator mode at a frequency $\omega_0$, and (\emph{b}) imperfections and nonlinearities in the cavity which can cause transfer of power from the spectator to the signal mode. At a more microphysics level this effect may be understood by the spectator mode driving some currents in a surface of the cavity which would overlap with the signal. Therefore, both of these effects can be studied by understanding the effect of a localized current within the cavity .
\begin{eqnarray}
(\partial^2_t-\vec{\nabla}^2)\vec{B}&=&\vec{\nabla}\times\vec{J}\nonumber\\
(\partial^2_t-\vec{\nabla}^2)\vec{E}&=&\partial_t\vec{J}
\eeq
In what follows we take a factorized form for the current $\vec{J}=\vec{j}(\vec{x})g(t)$. We now consider the two potential sources of leakage in turn.

\subsubsection{Driving Source leakage}
We now consider (\emph{a}), the driving source of the spectator mode.
To this end $\vec j$ will parametrize the location of the antenna and $g(t)\sim e^{i\omega_0 t}$ chosen to resonantly excite the mode $\mathbb{E}_0$
\beq \label{eq:E0}
\mathbb{E}_0\tilde{e}_0(\omega)=\frac{i\omega_0}{\omega^2-\omega_0^2-i\omega\omega_0/Q_0} \frac{\frac1{V}\int \vec{E}_0^*\cdot \vec{j}}{\sqrt {\frac1{V}\int |\vec{E}_0|^2}}\tilde{g}(\omega)\,,
\eeq
where
\beq\label{eq:eta0main}
\mathbb{E}_0\equiv \sqrt{\frac1V \int_V|\vec{E}_0(\vec{x})|^2}\equiv\eta_0 E_{\mathrm{peak}}\,.
\eeq
We assume that the source is a perfect sine wave of a quality that is higher than the cavity mode (with the understanding that this presents an experimental challenge). It is important to notice that in this case the frequency width of the field in Equation~(\ref{eq:E0}) is set by the source, and \emph{not} by the mode that is being populated. This is a consequence of a well known result, that a resonantly forced harmonic oscillator, at late times, will be excited at the frequency of the driving force, since oscillations at its eigenfrequency will damp away at late times.

The off-shell contribution to the signal from the spectator mode can be a source to the noise in the detector, which can be written as
\beq
\label{eq:E1}
\mathbb{E}_1\tilde{e}_1(\omega)=\frac{i\omega_0}{\omega^2-\omega_1^2-i\omega\omega_1/Q_1} \frac{\frac1{V}\int \vec{E}_1^*\cdot \vec{j}}{\sqrt {\frac1{V}\int |\vec{E}_1|^2}}\tilde{g}(\omega),
\eeq
where again, the spectrum of the driving source $\tilde g (\omega)\sim \delta(\omega-\omega_0)$ sets the field at late times. 
There are two ways to suppress this leakage contributions to the signal.
If we take a narrow signal bandwidth, effectively defining
\beq 
\label{eq:Sig}
S= \int_{\omega_1 -\Delta\omega_1}^{\omega_1 +\Delta\omega_1}  \mathbb{E}_1\tilde{e}_1(\omega) d\omega\,,
\eeq
in the limit of a pure delta function source, the contribution to $S$ from Equation~\eqref{eq:E1} vanishes.
The source will only contribute to the signal to the extent that the source width extends from $\omega_0$ to $\omega_1$. As stated above this can be parametrically much smaller than the cavity's $Q^{-1}$. The true signal, which is very narrow, will not be suppressed by this.

In addition to the suppression in the time domain, Equation~(\ref{eq:E1}) can also be suppressed by the spatial integral by choosing the source location wisely.
For example, if TM$_{010}$ is the spectator mode we wish to establish in the detection cavity, we can insert a probe of length $L$ along the $z-$axis:
\beq
\vec{j}=\hat{z} \delta(r) \frac I{2\pi r},
\eeq
such that $\mathbb{E}_0= I \frac{L}{V\eta_0}\frac{Q_0}{\omega_0}$, where $\eta_0$ is defined in Equation~\eqref{eq:eta0main}.
Since $\vec{j}$ is chosen along $z$, generating a longitudinal $E$ field, the spatial overlap of the current with TE$_{0m\ell}$  in Equation~\eqref{eq:E1} will vanish. Here again, the source will leak into the signal only due to the degree of inaccuracy in the source's placement.

\subsubsection{Leakage of Spectator to Signal mode due to Nonlinearities}

We now consider the second source of leakage which may be present due to impurities and nonlinear effects. For example, the spectator mode can have a spatial overlap with an impurity presumably on the cavity  wall, and cause a localized current $\vec J_\mathrm{im}(\vec x,t)$. If the coupling to the current is purely linear, the induced current will oscillate with the frequency of the spectator mode $\omega_0$, which is set by the spectrum of the source, Equation~(\ref{eq:E0}). The discussion of the previous subsection will apply in this case, and the signal~(\ref{eq:Sig}) is unaffected. 

It is likely, however, that the nonlinear effects will be present at some level. 
Mechanisms for generating nonlinearities may arise from impurities, material effects, such as brief transitions from super to normal conduction~\cite{nethercot_nonlinear_1963}, nonlinear Meissner currents~\cite{PhysRevB.51.16233, Dahm1999}, or from nonlinear response of impurities on the cavity wall. These effects depend highly on the material properties of the superconductor, such as the level of material disorder~\cite{PhysRevB.51.16233} and its purity.
In the presence of nonlinearities, at the classical level, modes that are integer multiples of $\omega_0$ will be induced as well. 
If there were more than one spectator mode, any sum and difference linear combination would be generated as well. If the signal mode were degenerate with a spectator mode or a multiple of it, e.g. as in~\cite{Bogorad:2019pbu}, such an effect can resonantly populate the signal.

In our setup, however, we assume $\omega_1$ is removed from $\omega_0$ or multiples of it.   In this case, the leakage again will be suppressed by the spectral overlap between the spectator mode (which is set by the source) and the signal mode. Employing an ultra narrow source may also help mitigating nonlinear leakage in our setup. Note, however, that if the source of nonlinearity are impurities, these may also  be dissipative to some degree, leading to a broadening of the spectrum of the generated currents. On general grounds we can limit the loss rate from the spectator mode to any source to $\omega_0/Q_0$. The signal mode, also by assumption, 
has a high $Q_1$ and thus its coupling to dissipative impurities is also suppressed. In this case a naive guess is that the leakage power would be suppressed by the product of the two quality factors. In addition, nonlinear transfer of power to the signal mode may also be suppressed by minimizing the spatial integral in Equation~(\ref{eq:E1})~\cite{Berlin:2019ahk}. Interestingly, a suppression factor of the coupling between two nearly degenerate modes has been achieved in some cases~\cite{Ballantini:2005am}. 

Given the dependence of these effects on the material science, an experimental and material theory effort is warranted. Nonlinear response may be studied for various material samples that with different fabrication techniques and with different surface treatments\footnote{Such studies have been initiated within the SQMS NQI center at Fermilab.}. The leakage effects in the receiver can also be studied in the experiment itself, in the absence of a powered emitter.

\section{Conclusion}\label{sec:conclusions}
In this paper, we considered a LSW experiment searching for axion-like particles using two identical SRF cavities. 
SRF cavity enables a great enhancement in the initial photon flux and final signal photon build-up, but comes with the downside that a large static magnetic field can no longer be used. To circumvent the problem, 
ALPs are produced via two cavity modes, with $\omega_a$ equal to the sum or difference of the frequencies of the two modes. In the receiver, ALPs can convert back to photons with frequency $|\omega_a\pm\omega_0|$ in the background of a cavity mode-0. If the photons' frequency matches that of another cavity mode, they can be produced resonantly. The frequency matching is guaranteed if the spectator in the receiver is chosen to be the same as one of the production modes in the emitter.

SRF cavities are already being used in a LSW setup to search for dark photons at Dark SRF~\cite{DarkSRF2}, in which several of the needed components of our proposal were demonstrated. These include high emitter powers, Hz level frequency control, and measurement protocols to avoid cross-talk.
In our setup, apart from the thermal background, a potential important source of noise may be the leakage to signal from the presence of a large amount of spectating photons in the receiver. As discussed in Section~4, the leakage can be attributed to an imperfect driving source of the spectator mode or impurities on the cavity wall, and can be effectively mitigated thanks to the $\mathcal{O}(\rm GHz)$ separation between the spectator and signal frequencies. An experimental study of such backgrounds is well motivated.
Assuming only thermal background, the sensitivity that can be potentially achieved by this setup is comparable to that projected by ALPSII, and can be further enhanced if high-level cavity modes can be used.

\section{Acknowledgement}
We thank Kevin Zhou and Asher Berlin for pointing out a mistake in the previous version of the paper. 
We thank Daniel Bafia, Anna Grasselino, Ryan Janish, Surjeet Rajendran, Alex Romananko, and especially Zhen Liu, Sam Posen, and Jim Sauls for useful discussions. CG is grateful to Yiming Zhong for his aid in data visualization. This paper is based on work supported by the U.S. Department of Energy, Office of Science, National Quantum Information (NQI) Science Research Centers through the Fermilab SQMS NQI Center. Fermilab is opertated by the Fermi Research Alliance, LLC under Contract DE-AC02-07CH11359 with the U.S. Dept. of Energy.

\pagebreak

\appendix

\section{Vacuum Modes of Cylindrical Cavity}\label{app:vacmodes}
The wave equation for $\vec{E}$ (and $\vec{B}$) in the vacuum is given by
\[
\vec{\nabla}^2\vec{E}-\frac1{c^2}\partial^2_t\vec{E}=0
\]
We want to work out the standing wave solutions in a cylindrical cavity of length $L$ and radius $R$.

\subsection{TM modes}
$B_z=0$. We assume that the form of $E_z$ is given by
\[
E_z=E_z(r)e^{in\theta}e^{ik_zz}e^{-i\omega t} 
\]
where $k_zL=\ell\pi$, and $n,\ell=0,1,2,\cdots$.
Hence,
\[
\frac 1r\partial_r(r\partial_rE_z)+\left(\omega^2-(\frac{\ell\pi}L)^2-(\frac nr)^2\right)E_z=0
\]
Requiring $E_z$ to be finite at $r=0$:
\[
E_{z}(r)=E_{0}J_n(r\sqrt{\omega^2-(\frac{\ell\pi}L)^2} )
\]
where $E_0$ is the field strength on the axis.
Requiring $E_z$ to vanish at the cavity wall $r=R$ picks out a discrete set of $\omega$ such that
\beq
\omega^{TM}_{nm\ell}=\sqrt{(\frac{Z_{nm}}R)^2+(\frac{\ell\pi}L)^2}
\eeq
where $Z_{nm}$ is the $m$th zero for the $n$th Bessel $J$.

One can obtain the rest of $\vec{E}$ and $\vec{B}$ via $\partial_t\vec{B}=-\vec{\nabla}\times\vec{E}$ and $\vec{\nabla}\cdot\vec{E}=0$. The mode that is of particular interest in this work is when $n=0$:
\beq
\vec{E}^{TM}_{0m\ell}(\vec{x},t)=E_0\left(
\begin{array}{c}
-i\frac{\ell\pi}L \frac{R}{Z_{0m}}J_1\left(r\frac{Z_{0m}}R\right)\\
0\\
J_0\left( r\frac{Z_{0m}}R\right)
\end{array}
\right)
e^{i\ell\pi z/L-i\omega^{TM}_{0m\ell}t}
\eeq
\beq
\vec{B}^{TM}_{0m\ell}(\vec{x},t)=B_0\left(
\begin{array}{c}
0\\
-i\omega^{TM}_{0m\ell} \frac{R}{Z_{0m}}J_1\left(r\frac{Z_{0m}}R\right)\\
0
\end{array}
\right)
e^{i\ell\pi z/L-i\omega^{TM}_{0m\ell}t}
\eeq

\subsection{TE modes}
$E_z$=0. We assume that the form of $B_z$ is given by
\[
B_z=B_0J_n(\sqrt{\omega^2-(\ell\pi /L)^2}r)e^{in\theta}e^{i\ell\pi z/L}e^{-i\omega t} 
\]
where $B_0$ is the field strength on the axis. Requiring that $B_z$ vanishes at the end caps:
\[
\ell=1,2,\cdots
\]
Imposing the boundary condition $\partial_rB_z|_R=0$:
\beq
\omega^{TE}_{nm\ell}=\sqrt{(\frac{S_{nm}}R)^2+(\frac{\ell\pi}L)^2}
\eeq
where $S_{nm}$ is the $m$th extremum of the $n$th Bessel $J$.
Again, it is the $n=0$ modes that we are interested in:
\beq
\vec{B}^{TE}_{0m\ell}(\vec{x},t)=B_0\left(
\begin{array}{c}
-i\frac{\ell\pi}L \frac{R}{S_{0m}}J_1\left(r\frac{S_{0m}}R\right)\\
0\\
J_0\left( r\frac{S_{0m}}R\right)
\end{array}
\right)
e^{i\ell\pi z/L-i\omega^{TE}_{0m\ell}t}
\eeq
\beq\label{eq:temode}
\vec{E}^{TE}_{0m\ell}(\vec{x},t)=B_0\left(
\begin{array}{c}
0\\
i\omega^{TE}_{0m\ell} \frac{R}{S_{0m}}J_1\left(r\frac{S_{0m}}R\right)\\
0
\end{array}
\right)
e^{i\ell\pi z/L-i\omega^{TE}_{0m\ell}t}
\eeq

\subsection{$E\cdot B$}
As shown above, $n>0$ introduces an $e^{in\theta}$ factor in the mode, which destroys the rotational symmetry of the system, thus not desirable.
With TE$_{0m'\ell'}$ and TM$_{0m\ell}$,
\beq
\begin{split}
(\vec{E}\cdot\vec{B})_{\omega_{\pm}}=&\frac{ E_{\rm peak} B_{\rm peak}}2 \Big(J_0(Z_{0m}r/R)J_0(S_{0m'}r/R)\\
&\pm
\frac{\omega^{0m\ell}_{\rm TM}\omega^{0m'\ell'}_{\rm TE}-k^{\ell}_zk^{\ell'}_z}{(Z_{0m}/R)(S_{0m'}/R)}J_1(Z_{0m}r/R)J_1(S_{0m'}r/R)\Big)
e^{i\pi (k^{\ell}_z\pm k^{\ell'}_z)z}
\end{split}
\eeq
where $k^{\ell}_z=\frac{\ell\pi}L$,
\beq
\omega_{\textrm{TM}}^{nm\ell}=\sqrt{(\frac{Z_{nm}}R)^2+(\frac{\ell\pi}L)^2},\quad\ell=0,1,2,\cdots,
\eeq
and
\beq
\omega_{\rm TE}^{nm\ell}=\sqrt{(\frac{S_{nm}}R)^2+(\frac{\ell\pi}L)^2},\quad\ell=1,2,\cdots.
\eeq
$Z_{nm}$ and $S_{nm}$ are the $m$th zero and the $m$th extremum of the $n$th Bessel function, respectively.

\section{Derivation of $P_{sig}$}\label{app:P_sig}
We closely follow the frequency conversion method introduced in \cite{Berlin:2019ahk}.

Let the signal mode be $
\vec{E}_{sig}(t,\vec{x})=\vec{E}_1(\vec{x})e_1(t)$, where $e_1(t)\sim e^{i\omega_1t}$,
and define the characteristic amplitude for the signal mode as
\beq\label{eq:eta1-app}
\mathbb{E}_1\equiv \sqrt{\frac1V \int_V|\vec{E}_1(\vec{x})|^2}
\eeq
Assume that the bandwidth of $\omega_1$ is given by $\omega_1/Q_1$, where $Q_1$ is the quality factor of the cavity. The steady state average power output can be expressed as
\beq
\begin{split}
P_{sig}=&\frac{\omega_1}{Q_1}\int_{V_{dc}}|\vec{E}_1(\vec{x})|^2\langle |e_1(t)|^2\rangle\\
=&\frac{\omega_1}{Q_1}V_{dc}\mathbb{E}_1^2\frac 1{(2\pi)^2}\int d\omega d\omega' \langle \tilde{e}_1(\omega)\tilde{e}^*_1(\omega')\rangle e^{i(\omega-\omega')t},
\end{split}
\eeq
where we have used Equation~\eqref{eq:eta1-app} and the Fourier transform of $e_n(t)$:
\beq
e_n(t)=\frac1{2\pi}\int d\omega e^{i\omega t}\tilde{e}_n(\omega),\quad
\tilde{e}_n(\omega)=\int dt e^{-i\omega t}e_n(t).
\eeq
The quantity $\langle\tilde{e}_1(\omega) \tilde{e}^*_1(\omega')\rangle$ can be interpreted as the power spectral density (PSD), and is expected to follow the relation
\beq\label{eq:psd}
\langle \tilde{e}(\omega) \tilde{e}^*(\omega')\rangle=S_e(\omega)\delta(\omega-\omega').
\eeq 
Therefore,
\beq
\begin{split}\label{eq:P_sig-app}
P_{sig}=\frac{\omega_1}{Q_1} V_{dc}\frac{ \mathbb{E}^2_1}{(2\pi)^2}\int d\omega \, \langle\tilde{e}_1(\omega) \tilde{e}^*_1(\omega)\rangle.
\end{split}
\eeq

To calculate $P_{sig}$, we need work out the amplitude of the signal mode, given a spectating cavity mode and an axion field.
Let the spectating mode be
$\vec{B}_{spe}(t,\vec{x})=\vec{B}_0(\vec{x}) b_0(t)$ and $\vec{E}_{spe}(t,\vec{x})=\vec{E}_0(\vec{x}) e_0(t)$, where $b_0(t)\sim e^{i\omega_0t}$ and $e_0(t)\sim -i e^{i\omega_0t}$.
Similarly, the characteristic amplitudes $\mathbb{E}_0(=\mathbb{B}_0)$ can be defined as
\beq\label{eq:eta0}
\mathbb{E}_0\equiv \sqrt{\frac1V \int_V|\vec{E}_0(\vec{x})|^2}\equiv\eta_0 E_{\mathrm{peak}}
\eeq
where $\eta_0\sim O(1)$. Recall that the laboratory produced axion field has spatial dependence as well as time dependence, hence $
a(\vec{x},t)\equiv \mathbf{a} (\vec{x})f(t)$, where $f(t)\sim e^{-i\omega_{a} t}$.
Starting with Equation~\eqref{eq:waveeqn}, we first write $\vec{E}$ on the l.h.s. as a linear combination of the vacuum cavity modes, $\vec{E}=\sum_n \vec{E}_n(\vec{x}) e_n(t)$. Therefore,
\beq
\begin{split}
-\sum_n\left(\omega_n^2\vec{E}_n+\frac{\omega_n}{Q_n}\partial_t\vec{E}_n+\partial_t^2\vec{E}_n\right)e_n(t)=-g\partial_t(\vec{E}_{spe}\times\vec{\nabla}a) 
+ g\partial_t(\vec{B}_{spe}\partial_ta)
\end{split}
\eeq
where a damping factor for each mode is inserted by hand.
Apply $\int dt\, e^{-i\omega t}$ to both sides and integrate by parts when necessary:
\beq\label{eq:master-app}
\begin{split}
\sum_n\left(\omega^2-\omega_n^2-i\frac{\omega\omega_n}{Q_n}\right)\vec{E}_n(\vec{x})\tilde{e}_n(\omega)
=\int dt\, e^{-i\omega t}\Big(-g\partial_t(\vec{E}_{spe}\times\vec{\nabla}a)
 + g\partial_t(\vec{B}_{spe}\partial_ta)
 \Big).
\end{split}
\eeq

Substituting $a,\vec{B}_{spe},\vec{E}_{spe}$ in terms of their spatial and temporal components, the r.h.s. of Equation~\eqref{eq:master-app} becomes
\beq
\begin{split}
-i\omega g\vec{E}_0\times\vec{\nabla}\mathbf{a} \int \frac{d\omega'}{2\pi}\tilde{e}_0(\omega-\omega')\tilde{f}(\omega')
+i\omega g\vec{B}_0\mathbf{a} \int d\omega' \frac {i\omega'}{2\pi} \tilde{b}_0(\omega-\omega') \tilde{f}(\omega')
\end{split}
\eeq
Since $e_0(t)$ and $b_0(t)$ are the same vacuum mode, they are related in a way that $-i b_0(t)=e_0(t)$, and so are their Fourier transforms.
Applying $\int_V \vec{E}^*_1$ to both sides of Equation~\eqref{eq:master-app}, followed by using the orthogonality relation between the vacuum modes on the l.h.s., Equation~\eqref{eq:master-app} becomes
\beq
\begin{split}\label{eq:sigE-app}
\mathbb{E}_1\tilde{e}_1(\omega)=&\frac{-i\omega g\mathbb{E}_0}{\omega^2-\omega_1^2-i\omega\omega_1/Q_1}
\times \int \frac{d\omega'}{2\pi}\tilde{e}_0(\omega-\omega')\tilde{f}(\omega')(\alpha +\beta\omega'+\frac1{\omega}\gamma),
\\
\mbox{where } &\alpha\equiv\frac{\int_V\vec{E}_1^*\cdot(\vec{E}_0\times\vec{\nabla}\mathbf{a}  )}{\sqrt{\int_V|\vec{E}_1|^2}\sqrt{\int_V|\vec{E}_0|^2}},\quad
\beta\equiv\frac{\int_V\vec{E}_1^*\cdot(\vec{B}_0\mathbf{a} )}{\sqrt{\int_V|\vec{E}_1|^2}\sqrt{\int_V|\vec{B}_0|^2}}.
\end{split}
\eeq
Now we are ready to compute $ \mathbb{E}_1^2\langle\tilde{e}_1(\omega)\tilde{e}^*_1(\omega)\rangle$, which is needed in computing $P_{sig}$.
\beq
 \mathbb{E}_1^2\langle\tilde{e}_1(\omega)\tilde{e}^*_1(\omega)\rangle= \frac{\omega^2g^2\mathbb{E}_0^2}{(\omega^2-\omega_1^2)^2+\omega^2\omega_1^2/Q_1^2}
\int\frac{d\omega'}{(2\pi)^2}S_{e_0}(\omega-\omega')S_f(\omega')\left|\alpha+\beta\omega'
\right|^2,
\eeq
where Equation~\eqref{eq:psd} is used in computing terms such as $\langle \tilde{f}(\omega)\tilde{f}^*(\omega')\rangle$.
Since $a$ has a negligible width, the PSD for $f$ is given by
$S_f(\omega)=\pi^2(\delta(\omega+\omega_{a})+\delta(\omega-\omega_{a}))$.
Finally, from Equation~\eqref{eq:P_sig-app}, we obtain an expression of the signal power:
\begin{equation}\label{eq:P_sig-general}
\begin{split}
P_{sig}=&\frac1{16\pi^2}\frac{\omega_1V_{dc}g^2\mathbb{E}_0^2}{Q_1}\int d\omega \frac{\omega^2}{(\omega^2-\omega_1^2)^2+\omega^2\omega_1^2/Q_1^2}\times\\
&\left\{
S_{e_0}(\omega+\omega_{a})|\alpha-\beta\omega_{a}
|^2
+S_{e_0}(\omega-\omega_{a})|\alpha+\beta\omega_{a}
|^2
\right\}\\
=&\frac1{16\pi^2}\frac{\omega_1V_{dc}g^2\mathbb{E}_0^2}{Q_1}\int _{\omega_-}^{\omega_+}d\omega \frac{\omega^2}{(\omega^2-\omega_1^2)^2+\omega^2\omega_1^2/Q_1^2}S_{e_0}(\omega-\omega_{a})
\left\{|\alpha-\beta\omega_{a}
|^2
+|\alpha+\beta\omega_{a}
|^2
\right\}
\end{split}
\end{equation}

Assuming that the spectating mode is perfectly peaked at one frequency with no width, i.e. $S_{e_0}(\omega)=\pi^2(\delta(\omega+\omega_0)+\delta(\omega-\omega_0))$,
\begin{equation}
P^{\rm ideal}_{sig}=\frac 18 V_{dc} \mathbb{E}_0^2 g^2\frac{Q_1}{\omega_1}\left(
|\alpha|^2+|\beta (\omega_1\mp\omega_0)
|^2
\right),
\end{equation}
where in the last line, to get the signal on resonance, we used frequency matching condition:
\beq
\omega_1\mp\omega_0=\omega_{a}.
\eeq

\bibliographystyle{unsrt}
\bibliography{ref}

\end{document}